\newcommand{\myfig}[5]{
\begin{figure}[{#5}]
\includegraphics[keepaspectratio,width=#4,angle=0]{#1}%
\caption{#2}\label{#3}%
\end{figure}}

\newcommand{\secref}[1]{Section\ \ref{#1}}

\documentclass[12pt,aps,pre,amsmath,amssymb,hidelinks]{article}
\usepackage{authblk}
\usepackage{setspace}
\usepackage[margin=0.5in]{geometry}

\usepackage{amsmath}
\usepackage{amsthm}

\usepackage[bookmarks=false]{hyperref}

\hypersetup{
  colorlinks=false,
  pdfborder={0 0 0}
}

\usepackage{eucal}
\usepackage{amssymb}
\usepackage{mathrsfs}
\usepackage{hyperref}

\usepackage{graphicx,color,subfigure}

\title{Automatic Segmentation of Fluorescence Lifetime Microscopy Images of Cells Using Multi-Resolution Community Detection}

\renewcommand\footnotemark{}
\renewcommand\footnoterule{}

\author[1]{Dandan Hu$^{*}$}
\author[2]{Pinaki Sarder$^{*}$}
\author[1]{Peter Ronhovde}
\author[3]{Sandra Orthaus}
\author[2]{Samuel Achilefu}
\author[1]{Zohar Nussinov$^{\mathrm{\dagger}}$}
\affil[1]{Department of Physics, Washington University, One
Brookings Drive, Campus Box 1105, St. Louis, MO~63130}
\affil[2]{Department of Radiology, Washington University School of
Medicine, 4525 Scott Avenue, Campus Box 8225, St. Louis, MO~63110}
\affil[3]{PicoQuant GmbH, Berlin, Germany}

\begin{document}

  \maketitle
  \footnoterule{\hspace{-3.5ex}$^{*}$Dandan Hu and Pinaki Sarder contributed equally to this work.\\
								$^{\mathrm{\dagger}}$Corresponding author. Email: zohar@wuphys.wustl.edu, Ph: 314-935-6277, Fax: 314-935-6219.}

 \newpage

\begin{abstract}
We have developed an automatic method for segmenting fluorescence lifetime (FLT) imaging microscopy (FLIM) images of cells inspired by a multi-resolution community detection (MCD) based network segmentation method. The image processing problem is framed as identifying segments with respective average FLTs against a background in FLIM images. The proposed method segments a FLIM image for a given resolution of the network composed using image pixels as the nodes and similarity between the pixels as the edges. In the resulting segmentation, low network resolution leads to larger segments and high network resolution leads to smaller segments. Further, the mean-square error (MSE) in estimating the FLT segments in a FLIM image using the proposed method was found to be consistently decreasing with increasing resolution of the corresponding network. The proposed MCD method outperformed a popular spectral clustering based method in performing FLIM image segmentation. The spectral segmentation method introduced noisy segments in its output at high resolution. It was unable to offer a consistent decrease in MSE with increasing resolution.

\end{abstract}


\newpage

\section{Introduction} \label{sec:glassesintroduction}
Fluorescence lifetime imaging microscopy (FLIM), a promising technique for imaging molecular process, generates images using the characteristic fluorescence lifetimes (FLTs) from sub-cellular locations of biological samples (such as cells and thin tissue sections) that are treated with fluorescent contrast agents. The FLT is the average time a molecule resides in the excited state before returning to the ground state through fluorescence emission (Nothdurft {\em et al.}, 2012). In this work, we propose a multi-resolution community detection (MCD) method based on graph partitioning theory (Fortunato, 2010) to automatically segment FLIM images of cells. MCD (Fortunato, 2010; Girvan \& Newman, 2002; Hu {\em et al.}, 2012;  Newman, 2004; Ronhovde \& Nussinov, 2009; Ronhovde \& Nussinov, 2010) seeks to divide groups of nodes with dense connections internally and with sparser connections between the groups in a network. It thus partitions a large physically interacting system into optimally decoupled communities. To demonstrate the performance of the proposed method, we segmented donar FLTs in FLIM images of cells transfected with EGFP-RFP fusion F$\ddot{\mathrm{o}}$rster resonance energy transfer (FRET) protein pairs (Orthaus {\em et al.}, 2009).  

Image segmentation plays a crucial role in medical imaging applications by enhancing the detection of biological structures of interest. Existing medical image segmentation methods (Pham {\em et al.}, 1998) are typically based on spectral clustering, normalized cuts (Fortunato, 2010; Ng {\em et al.}, 2002; Perona \& Freeman, 1998; Scott \& Longuet-Higgins, 1990; Shi \& Malik, 2000), and mixture of Gaussian distributions (MGD) (Dempster {\em et al.}, 1977). FLIM is a relatively newer technique to the medical imaging community. Existing FLIM image analysis software packages (SPCImage, Becker-Hickl, Germany; SymPhoTime, PicoQuant, Germany; VistaVision, ISS Inc., Champaign, IL) deliver FLIM images and corresponding FLT histograms for FLIM system acquired data. VistaVision software package further provides phasor histograms of FLIM images (Stringaria {\em et al.}, 2011). Users are able to manually segment pixels corresponding to distinct FLTs in FLIM images based on their locations in the FLT histograms or the phasor histograms. However, no established method is available for automatically segmenting FLIM images in the literature.

The proposed MCD method performs automatic unsupervised segmentation of FLIM images for a given resolution of the network composed using image pixels as the nodes and similarity between the pixels as the edges. During this process, the input images are segmented starting from different initial states, and significant segments are determined in the final segmented images using information theoretic correlations. Low network resolution leads to larger segments and high network resolution leads to smaller segments. The outcome is a segmented image containing distinct average FLTs in each of its segments.

We compared the performance of the proposed method with a popular spectral clustering method developed by Ng {\em et al.} in segmenting FLIM images. The mean-square error (MSE) in estimating the FLT segments in a FLIM image using the MCD method was found to be consistently decreasing with increasing resolution of the network constructed using the FLIM image. In contrast, the spectral clustering was unable to deliver a decrease in MSE in segmenting FLIM images with increasing resolution, and this method introduced noisy segments in its output at high resolution. 

The study is presented as follows. In \secref{sec:nearinfared}, FLIM imaging and its applications are discussed. In \secref{sec:2Dmethod}, the proposed MCD method for FLIM image segmentation is described. In \secref{sec:results}, the performance of the proposed method in segmenting FLIM images of cells transfected with EGFP-RFP fusion FRET protein pairs is demonstrated, and this performance is compared with that attained using the spectral clustering method developed by Ng {\em et al.} We conclude in \secref{sec:conclusion}. 

\section{Fluorescence Lifetime Imaging Microscopy}\label{sec:nearinfared}

FLIM is a promising technique for imaging molecular processes. FLIM is typically performed in the frequency or time domain. In the frequency domain, a sinusoidal modulated ($0.1$-$1$ GHz) light source illuminates the sample, and FLTs are measured by detecting and analyzing the amplitude and phase shift between the excitation light and fluorescence emission (Gadella {\em et al.}, 1993). In the time domain, pulsed light illuminates the sample, and the time-course of fluorescence emission is detected and analyzed for FLTs. Imaging systems use either time-gated wide field image intensifiers (Elson {\em et al.}, 2002) or time-resolved laser scanning point detection (Morgan {\em et al.}, 1995). FLIM images depict FLTs of fluorophore molecules in each pixel corresponding to the sample micro-environment. Applications of FLIM include imaging molecular signalling (Webb {\em et al.}, 2008) and trafficking (Verveer {\em et al.}, 2000), imaging spatial concentration of intracellular ions (Lahn {\em et al.}, 2011), assessing intracellular environment (Kneen {\em et al.}, 1998), characterizing tissue slices {\em in vivo} (Ushakov {\em et al.}, 2011), and determining molecular interactions using FRET (Keese {\em et al.}, 2010). Existing literature does not enlist any automatic method for segmenting FLIM images. This void leaves the users to manually select regions with distinct FLTs in the FLIM images based on FLT histograms or phasor histograms computed by existing FLIM image analysis software packages. Adding a feature in these packages for automatic segmentation of FLIM images will eliminate further the manual intervention in analyzing FLIM data using these software packages, and it would thus be useful for the community. The work presented herein seeks to contribute in this direction. 

\section{Segmentation Using Multi-Resolution Community Detection}\label{sec:2Dmethod}

\subsection{Potts Model Hamiltonian}
To segment a FLIM image, we construct a network by using the image pixels as nodes and the absolute FLT difference between two pixels as the weight between the nodes formed by these pixels. The MCD method segments the nodes of the resulting network, by minimizing a Potts model Hamiltonian,
\begin{eqnarray}
{\cal{H}} = \frac{1}{2}\sum_{i\not=j}
(W_{ij}-\overline{W})\Big[\Theta(\overline{W}-W_{ij})+\gamma\Theta(W_{ij}-\overline{W})\Big]
\delta(\sigma_{i}, \sigma_{j}).
 \label{eq:newpotts}
\end{eqnarray}
The weight $W_{ij}$ denotes the absolute FLT difference between a pixel pair formed by the $i^{\mathrm{th}}$ and $j^{\mathrm{th}}$ $(\{i,j\}\in\{1,2,\dots,N\})$ pixels in the input image with $N$ pixels, and $\overline{W}$ denotes the background of $W_{ij}.$ The Heaviside function $\Theta(\cdot)$ ``turns on'' or ``off'' the edge designation. 
\begin{eqnarray}
\Theta(W_{ij}-\overline{W})=\begin{cases}1, &\quad\text{if
$W_{ij}>\overline{W}$},\\0, &\quad\text{otherwise} .
\end{cases}\label{eq:theta}
\end{eqnarray}
The parameter $\gamma$ controls the resolution of the estimated segments. Decreasing $\gamma$, the minima of Eq.\ (\ref{eq:newpotts}) lead to solutions progressively lower intra-community edge densities, effectively ``zooming out'' toward larger segments. The Kronecker delta $\delta(\cdot)$ is given by,
\begin{eqnarray}
\delta(\sigma_i,\sigma_j)=\begin{cases}1, &\quad\text{if
$\sigma_i=\sigma_j$},\\0,&\quad\text{otherwise}.
\end{cases}\label{eq:delta}
\end{eqnarray}

In the above Hamiltonian, by virtue of the $\delta(\sigma_{i}, \sigma_{j})$ term, each spin $\sigma_i$ interacts only with other spins in its own segment. The spin $\sigma_i$ $(\forall\sigma_i\in\{1,2,\dots,K\})$ defines the segment identity for the $i^{\mathrm{th}}$ $(i\in\{1,2,\dots,N\})$ pixel, and the algorithm optimizes it by minimizing the energy defined by Eq.\ (\ref{eq:newpotts}). As such, the resulting model is {\em local}---a feature that enables high accuracy along with rapid convergence (Ronhovde \& Nussinov, 2010). Thus, minimizing the Hamiltonian of Eq.\ (\ref{eq:newpotts}) corresponds to identifying strongly connected segments of pixels.

\subsection{Community Detection}\label{sec:CD} 

The community detection (CD) algorithm minimizes Eq.\ (\ref{eq:newpotts}) using four steps (Ronhovde \& Nussinov, 2010).

\begin{enumerate}

\item The pixels are partitioned based on a symmetric or fixed $K$ initialization.

\begin{itemize}
\item Symmetric initialization is used for the unsupervised case, where each pixel forms its own segment; i.e., initially, there are $K^{(0)}=N$ segments. Here the algorithm does not know what the number of segments are, so the symmetric initialization provides the advantage of no bias towards a particular segment. The algorithm decides the number of segments $K,$ by means of the lowest energy solution. In the current work, we perform such unsupervised image segmentation.
\item Fixed $K$ initialization is used in a supervised image segmentation, where all pixels are divided into $K$ segments using a random initial distribution. The community membership of an individual pixel is then changed to lower the solution energy using CD algorithm. Here the user decides about the number of initial segments $K$ based on the desired information. For instance, if only one target needs to be identified, $K=2$ is enough, which describe the target and background.
\end{itemize}

\item Each pixel is then placed in the segment that best lowers the energy of Eq.\ (\ref{eq:newpotts}) based on the current state of the system.

\item This process is repeated for all pixels. The iteration is continued until no energy lowering moves are found after one full cycle through all pixels.

\item The above three steps are repeated for $T$ trials, and the lowest energy is selected as the best solution. Different trials differ solely by the permuted pixel order of the initial state. 

\end{enumerate}

\subsection{Multi-Resolution Community Detection}\label{sec:MCD} 

We illustrate below how the multi-resolution CD (MCD) algorithm (Ronhovde \& Nussinov, 2009) works.

To begin with the MCD algorithm, users first specify the number of replicas $R$ at each resolution $\gamma$, the number of trials $T$ per replica, and the starting and ending resolutions, $\gamma_0$ and $\gamma_f,$ respectively. See Appendix \ref{app:replica} for the definitions on ``trial'' and ``replica.'' We typically use $8\leq R\leq 12$ and $2\leq T\leq 20$. In the case of symmetric initialized state of one pixel per community, the initial state of the replicas are generated by permuting the pixel labels. These permutations $P$ simply reorder the pixel indices $(1,2,3,\dots,i,\dots, N) \to (P1,P2,\dots,PN)$ (with $Pi$ the state of $i$ under a permutation), and thus lead to a different initial state.  

\begin{enumerate}

\item The algorithm starts from the initialization of the system, as described in item (1) of Section \ref{sec:CD}.

\item Eq.\ (\ref{eq:newpotts}) is then minimized independently for all replicas at a resolution $\gamma = \gamma_{i} \in \{\gamma_{0},\gamma_{1},\dots,\gamma_{f-1},\gamma_{f}\},$ as described in Section \ref{sec:CD}.

\item The algorithm then calculates the average inter-replica information theoretic measures, such as $I_\mathrm{N}$ and $V,$ at each value of resolution $\gamma$ for the entire range
of the resolutions studied. Values of $\gamma$ corresponding to the extrema in the average inter-replica information theoretic overlaps results image segmentation that is locally insensitive to the change of resolution (i.e., small changes in $\gamma$) and generally highlights prominent features of the image. Different levels of detail and resolutions can be determined by setting the resolution parameter $\gamma$. 

\end{enumerate}

\section{Results}\label{sec:results}

This section describes representative examples of the proposed MCD method using FLIM images of cells transfected with EGFP-RFP fusion FRET protein pairs (Orthaus {\em et al.}, 2009). A performance comparison between the MCD and a popular spectral clustering method developed by Ng {\em et al.} in segmenting FLIM images is also discussed.

\subsection{Datasets} 
To demonstrate the performance of the proposed MCD method, we employed two FLIM images of live 12V HC Red cells expressing a protein fused to EGFP (donor) and RFP (acceptor) separated by a short linker. The images were acquired using an Olympus FV1000 laser scanning microscope (LSM) equipped with the PicoQuant LSM Upgrade Kit for FLIM. Such a donor-acceptor fusion serves as a positive control for FRET. The samples were excited by pulsed excitation at 470 nm with 40 MHz repetition. Photons were detected by a single channel SPAD (PicoQuant, Germany) set-up. A fluorescence bandpass filter (500-540 nm) limited the detection to the donor (EGFP) fluorescence only.

The FLIM image in Fig.\ 1A shows two cells with different average donor FLTs: a FRET cell and a cell where the acceptor RFP was irreversibly bleached leading to a FLT shift from approximately 2.1 ns towards 2.4 ns. In Fig.\ 1A, by carefully adjusting the colormap of the image, this shift of the average FLT was clearly distinguished using yellow and red colors, respectively. In the FRET cell (shown using yellow color), half of the EGFP-RFP fusion proteins could adopt a proper conformation due to complete maturation allowing for FRET (Orthaus {\em et al.}, 2009). When analyzing the acceptor-bleached cell (shown using red color), the situation was different. Only 15\% of the EGFP molecules were quenched by energy transfer to some remaining acceptor molecules, whereas the majority of donor molecules (85\%) could not undergo FRET any more because an appropriate acceptor molecule was missing. In another similar image in Fig.\ 1B, an average donar FLT of 2.2 ns (shown using yellow) could be obtained in the cell corresponding to the quenched EGFP, and an average donar FLT of 2.9 ns (shown using red) was found in the other cell where the acceptor molecules were irreversibly destroyed. 

\subsection{Multiresolution Community Detection for Varying Resolution}

For the FLIM images shown in Fig.\ 1, we define the edge weight between two pixels as the absolute FLT difference between them. The MCD was applied to segment the resulting networks formed by the image pixels as nodes. Figs.\ 2G and 3G show the plots of the respective information theoretic overlaps between the replicas of the MCD, such as their normalized mutual information $I_\mathrm{N}$ and variation of information $V$, together with the respective number of estimated segments $\widehat{K}$ as a function of the resolution parameter $\gamma$. Decreasing $\gamma$, the minima of Eq.\ (\ref{eq:newpotts}) leads to solutions with progressively lower intra-segment edge densities, effectively ``zooming out'' toward larger segments. Natural network resolutions correspond to the values of $\gamma$ for which the replicas exhibit extrema and plateau in the average of their information theoretic overlaps when expressed as a function of $\gamma$ (Ronhovde \& Nussinov, 2009). Recall that the independent solutions of the MCD method attained from different starting points are defined as replicas in Section \ref{sec:MCD}.

Figs.\ 2A-2F and 3A-3F show the results of the automatic image segmentation using our MCD algorithm at different resolutions for the two FLIM images shown in Fig.\ 1. The segments are depicted using false colors. As the resolution increases from Fig.\ 2A to 2F and from Fig.\ 3A to 3F, the images show more detailed segments. In Figs.\ 2D-2F and 3D-3F, two major segments, one representing the respective FRET cell and the other representing the respective unquenched cell, are clearly visible for $\gamma > 1$, in addition to the respective background. Thus, by using different resolutions $\gamma$, the MCD method was able to detect the segments at different scales. To generate the segmented images, the number of replicas used was 8 and the number of trials used was 1. Implementation of the automatic segmentation parallely in different resolutions will allow users to obtain segments without having any human intervention of adjusting the image colormap.   

\subsection{Spectral Clustering for Varying Resolution}

We compared the performance of the proposed MCD method in automatically segmenting FLIM images using a popular spectral clustering method developed by Ng {\em et al.} In brief, the adoption of this method in this paper first constructs a network with image pixels as nodes and edge weight between two nodes as the squared distance between the FLTs of the corresponding pixels smoothed by a Gaussian kernel. The affinity matrix formed by the resulting edge weights is then normalized, and eigen decomposition of the resulting matrix is performed. Eigen vectors corresponding to the eigen values $\lambda \geq \alpha$ are chosen, these eigen vectors are normalized again, and are segmented along the rows to segment the network. Similar to the MCD method, $\alpha$ determines the resolution of the segmented images. Increasing $\alpha$ restricts to confine with eigen vectors of smaller variation across the pixels, and decreasing $\alpha$ allows to include eigen vectors of larger variation across the pixels. Consequently, increasing $\alpha$ effectively leads to ``zoom out'' toward larger segments.   

Figs.\ 4A-4F and 4G-4L show the performance of the spectral clustering method in segmenting the FLIM images shown in Fig.\ 1 for decreasing $\alpha$. For the FLIM image shown in Fig.\ 1A, the segmented images shown in Figs.\ 4A-4F using false colors depict increasing noise at high resolution, and none of these segmented images clearly depicts the two major expected segments. For the FLIM image shown in Fig.\ 1B, using false colors, the segmented images show the two respective cells as one segment in Figs.\ 4G and 4H in low resolution and as two expected segments in Figs.\ 4I and 4J in a slightly higher resolution. At the resolution limit, similar to the case of Fig.\ 1, the segmented images for Fig.\ 2 were introduced with high amount of noise. In comparison to the proposed MCD method, spectral clustering at high resolution thus (i) was not able to provide the expected segments, and (ii) introduced high amount of noise in the segmented images. 

\subsection{Performance Comparison Between Multi-Resolution Community Detection and Spectral Clustering}

To quantitatively compare the MCD and spectral clustering methods, we compared mean-square errors (MSEs) in segmenting the major segments of the FLIM images shown in Fig.\ 1. To perform a fair comparison, following procedure was followed. The ground-truth FLTs of the cells in each FLIM image shown in Figs. 1A and 1B were computed to be the average FLTs in the yellow and red colored regions. The segmented images using the MCD method in Figs.\ 2A-2C and 3A-3C and those using the spectral clustering method in Figs.\ 4A-4D, 4F, and 4G-4H depict the two respective expected major segments as one segment. Consequently, average FLT of this single segment was used as the estimated FLT for both respective cells in each segmented image. The segmented images using the MCD method in Figs.\ 2D-2F and 3D-2F and those using the spectral clustering method in Fig.\ 4E and 4I-4L depict the two respective expected major segments. Consequently, average FLTs in these segments were used as the estimated FLTs for the two cells in each segmented image. The MSE in segmenting the FLIM images shown in Fig.\ 1 at each resolution was the squared distance between the estimated FLTs and the ground-truth FLTs of the respective cells. 

The MSE in estimating average FLTs of the correct segments using the MCD method consistently decreases with increasing resolution; see Fig.\ 5. The MCD method offers lower MSE than the spectral clustering method in its all network resolution for the FLIM image shown in Fig.\ 1A, and in its high network resolution region ($\gamma > ~10$) for the FLIM image shown in Fig.\ 1B; see Figs.\ 5-6. The MSE in estimating average FLTs of the correct segments using the spectral clustering method does not consistently decrease with increasing resolution for the FLIM image shown in Fig.\ 1A. For Fig.\ 1B, the MSE in estimating average FLTs of the correct segments using the spectral clustering method shows a decrease in increasing the resolution by decreasing $\alpha$ up to 0.0625. Decreasing $\alpha$ below this value introduces noisy segments in the output, and the MSE in estimating average FLTs of the correct segments becomes very high. Consequently, such MSEs at limiting resolutions are not shown in Fig.\ 6B for clarity. In summary, the proposed MCD method outperforms the spectral clustering method in MSE sense in automatically segmenting the FLIM images shown in Fig.\ 1. 

\section{Conclusion}\label{sec:conclusion}
We have developed a multiresolution community detection (MCD) algorithm to automatically segment fluorescence lifetime imaging microscopy (FLIM) data. The proposed method is able to identify segments in different scales in the input FLIM images. It outperforms a popular graph-based spectral clustering method developed by Ng {\em et al.} in segmenting FLIM images. The MCD method was able to provide correct segments for FLIM images of cells transfected with EGFP-RFP fusion FRET protein pairs. The spectral clustering method was unable to provide such correct segments and introduced high amount of noise in the segmented images at high resolution. The MCD method offers lower mean-square errors in segmenting the FLIM images with that obtained using the spectral clustering method. 

The MCD method for automatically segmenting FLIM images will allow to avoid any manual selection of regions with distinct FLTs in the FLIM images based on FLT histograms or phasor histograms computed by existing FLIM image analysis software packages. Adding a feature in these packages for automatic segmentation of FLIM images will thus minimize error in analyzing FLIM data using them.

The success of our bare MCD graph theory based method naturally suggests the possibility of yet more potent approaches which build on it. We briefly propose and speculate on a possible extension involving the use of known prior information. We hope to explore this possibility in future work. We may employ expectation maximization (EM; Dempster {\em et al.}, 1977) to a given image vis a vis a library of finite number of images of known tissue types to infer probabilities that different parts of the image will be locally similar to any of the previously known types. We may then use the similarity of these local probabilities to define weights in the graph and employ MCD. In this approach, the MCD will not invoke bare weights resulting from only local intensity strengths in an image (as we have in the current work). Rather, the MCD will use edge weights as given by these probabilities (inferred via EM).

\bigskip

\section*{Acknowledgement}
This work is supported by the National Science Foundation (NSF) under grant numbers NSF DMR-1106293, NSF 1066293 (Aspen Center for Physics) and by the National Institute of Health (NIH) under grant numbers NIH R01 EB008111, NIH R01 EB007276, NIH R33 CA123537,
NIH U54 CA136398, and NIH HHSN268207000046C. We wish to acknowledge a discussion with Prof.\ Veit Elser from Cornell University. The images shown in Fig.\ 1 were taken from the application note written by Orthaus {\em et al.}, with kind permission from PicoQuant.

\appendix

\section*{Appendices}

\section{Definitions: Trials and Replicas}\label{app:replica}
We review the notions of ``trials'' and ``replicas,'' used in our community detection (CD) algorithms. Both of these notions pertain to the use of multiple identical copies of the same system which differ from one another by a permutation of the initial site
indices. Thus, whenever the time evolution depends on sequentially ordered searches for energy lowering moves (as it does in our greedy algorithm), these copies may generally reach different local solutions. By the use of an ensemble of such identical copies, accurate results are attained as well as information theoretic correlations are determined (Appendix B) between the candidate solutions, and a detailed picture of the system is inferred from them.

In the definitions of ``trials'' and ``replicas'' given below, any given algorithm may be used to minimize the selected cost function. In our particular case, the Hamiltonian of Eq.\ \ref{eq:newpotts} is minimized.

$\bullet$ {\underline{{\em Trials:}} We use ``trials'' alone in our bare community detection algorithm. The algorithm is evaluated on the same problem $T$ independent times. This may generally lead to different contending states that minimize Eq.\ \ref{eq:newpotts}. Out of these $T$ trials, the lowest energy state is picked and that state is used as the solution.

$\bullet$ {\underline{{\em Replicas:}}  We use both ``trials'' and ``replicas'' in our MCD algorithm. Each sequence of the above described $T$ trials is termed as a replica.  When using ``replicas'' in the current context, the aforementioned $T$ trials (and pick the solution that attains lowest energy in the Hamiltonian of Eq.\ \ref{eq:newpotts}) are evaluated $R$ independent times. By examining information theoretic correlations between the $R$ replicas, we infer which features of the contending solutions are well agreed on (and thus are likely to be correct), and on which features there is a large variance between the disparate contending solutions that may generally mark important physical boundaries. The information theoretic correlations are computed within the ensemble
of $R$ replicas. Specifically, the information theoretic extrema as a function of the resolution parameter, generally correspond to more pertinent solutions that are locally stable to a continuous change of scale. It is in this way the important physical scales
in the system are detected.

\section{Information Theoretic Measures}
\label{app:information}

We use information theoretic measures to calculate correlations
between community detection (CD) solutions. The CD method partitions
$N$ pixels for a replica $r$ $( \forall r\in\{1,2,\dots,R\})$ into
$K_r$ segments, where segment $k$ ($k\in\{1,2,\dots,K_r\}$) consists
of $N_k$ pixels. The ratio $N_k/N$ is the probability that a
randomly selected pixel is found in the segment $k$
($k\in\{1,2,\dots,K_r\}$). 

The Shannon entropy (Hu {\em et al.}, 2012) is
\begin{eqnarray}
  H_r = -\sum_{k=1}^{K_r} \frac{N_k}{N}\log_2\frac{N_k}{N}.
  \label{eq:HA}
\end{eqnarray}

The mutual information $I(r,s)$ between the replicas $r$ and $s$
($\{r,s\}\in\{1,2,\dots,R\}$) is
\begin{eqnarray}
  I(r,s)=\sum_{k_1=1}^{K_\mathrm{r}}\sum_{k_2=1}^{K_\mathrm{s}}\frac{N_{k_1k_2}}{N}\log_2\frac{n_{k_1k_2}N}{n_{k_1}n_{k_2}},
  \label{eq:IAB}
\end{eqnarray}
where $N_{k_1k_2}$ is the number of common pixels in the segment
$k_1$ ($k_1\in\{1,2,\dots,K_\mathrm{r}\}$) of replica $r$
($r\in\{1,2,\dots,R\}$) and the segment $k_2$
($k_2\in\{1,2,\dots,K_\mathrm{s}\}$) of replica $s$
($s\in\{1,2,\dots,R\}$).

The variation of information $V(r,s)$ between the two segments $r$
and $s$ is
\begin{eqnarray}
  V(r,s)=H_r+H_s-2I(r,s),
\end{eqnarray}
which has a range of $0\leq V(r,s)\leq\log_2 N$.

The normalized mutual information $I_\mathrm{N}(r,s)$ is
\begin{eqnarray}
  I_\mathrm{N}(r,s)=\frac{2I(r,s)}{H_r+H_s},
\end{eqnarray}
with the obvious range of $0\leq I_\mathrm{N}(r,s)\leq 1$.

Higher $I_\mathrm{N}(\cdot)$ and lower $V(\cdot)$ values indicate
better agreement between the compared segments.

\newpage

\section*{References}
{}[Dempster {\em et al.}, 1977] Dempster, A., Laird, N., \& Rubin, D.\ (1977). Maximum likelihood from incomplete data via the expectation maximization algorithm. {\em Journal of the Royal Statistical Society.} {\bf 39}(1), 1--38.\\ 
{}[Elson {\em et al.}, 2002] Elson, D.S., Siegel, J., Webb, S.E.D., {\em et al.} (2002). Wide-field fluorescence lifetime imaging with optical sectioning and spectral resolution applied to biological samples. {\em Journal of Modern Optics.} {\bf 49}(5-6), 985--995.\\
{}[Fortunato, 2010] Fortunato, S.\ (2010). Community detection in graphs. {\em Physics Reports.} {\bf 486}(3-5), 75--174.\\
{}[Gadella {\em et al.}, 1993] Gadella, T.W.J., Jovin, T.M., \& Clegg, R.M. (1993). Fluorescence lifetime imaging microscopy (FLIM)--spatial-resolution of microstructures on the nanosecond time-scale. {\em Biophysical Chemistry.} {\bf 48}(2), 221--239.\\
{}[Girvan \& Newman, 2002] Girvan, M.\ \& Newman,  M.E.J.\ (2002). Community structure in social and biological networks. {\em Proceedings of the National Academy of Sciences of the USA.} {\bf 99}(12), 7821--7826.\\
{}[Hu {\em et al.}, 2012] Hu, D., Ronhovde, P., \& Nussinov, Z.\ (2012). Phase transition in random Potts systems and the community detection problem: Spin-glass type and dynamic perspectives. {\em Philosophical Magazine.} {\bf 92}(4), 406--445.\\
{}[Hu {\em et al.}, 2012] Hu, D., Ronhovde, P., \& Nussinov, Z.\ (2012). Replica inference approach to unsupervised multiscale image segmentation. {\em Physical Review E.} {\bf 85}(1 Pt 2), 016101.\\
{}[Keese {\em et al.}, 2010] Keese, M., Yagublu, V., Schwenke, K., Post, S., \& Bastiaens, P.\ (2010). Fluorescence lifetime imaging microscopy of chemotherapy-induced apoptosis resistance in a syngenic mouse tumor model. {\em International Journal of Cancer.} {\bf 126}(1), 104--113.\\
{}[Kneen {\em et al.}, 1998] Kneen, M., Farinas, J., Li, Y., \& Verkman, A.S.\ (1998). Green fluorescent protein as a noninvasive intra-cellular pH indicator. {\em Biophysical Journal.} {\bf 74}(3), 1591--1599.\\
{}[Lahn {\em et al.}, 2012] Lahn, M., Dosche, C., \& Hille, C.\ (2011). Two-photon microscopy and fluorescence lifetime imaging reveal stimulus-induced intracellular Na+ and Cl- changes in cockroach salivary acinar cells. {\em American Journal of Physiology: Cell Physiology.} {\bf 300}(6), C1323--C1336.\\
{}[Morgan {\em et al.}, 1995] Morgan, C.G., Murray, J.G., \& Mitchell, A.C.\ (1995). Photon-correlation system for fluorescence lifetime measurements. {\em Review of Scientific Instruments.} {\bf 66}, 3744--3749.\\
{}[Newman, 2004] Newman, M.E.J.\ (2004). Detecting community structure in networks. {\em European Physical Journal B.} {\bf 38}(2) 321--330. \\
{}[Ng {\em et al.}, 2002] Ng, A., Jordan, M., \& Weiss, Y. (2002). On spectral clustering: analysis and an algorithm. In Dietterich, T., Becker, S., and Ghahramani, Z., (Eds.), {\em Advances in Neural Information Processing Systems.} MIT Press. {\bf 14}, 849--856.\\
{}[Nothdurft {\em et al.}, 2012] Nothdurft, R., Sarder, P., Bloch, S., Culver, J., \& Achilefu, S.\ (2012). Fluorescence lifetime imaging microscopy using near-infrared contrast agents. {\em Journal of Microscopy.} {\bf 247}(2), 202--207.\\
{}[Orthaus {\em et al.}, 2009] Orthaus, S., Buschmann, V., B$\ddot{\mathrm{u}}$lter, A., {\em et al.} (2009). Quantitative {\em in-vivo} imaging of molecular distances using FLIM-FRET. Application note by PicoQuant GmbH, Berlin, Germany. 1--7.\\
{}[Perona \& Freeman, 1998] Perona, P.\ \& Freeman, W.\ T.\ (1998). A factorization approach to grouping. In Burkardt, H.\ \& Neumann, B.\ (Eds.), {\em Proceedings of European Conference on Computer Vision.} 655--670.\\
{}[Pham {\em et al.}, 1998] Pham, D.L., Xu, C., \& Prince, J.L.\ (1998). A survey of current methods in medical image segmentation. {\em Annual Review of Biomedical Engineering.} {\bf 2}, 315--337.\\
{}[Ronhovde \& Nussinov, 2009] Ronhovde, P.\ \& Nussinov, Z.\ (2009). Multiresolution community detection for megascale networks by information-based replica correlations. {\em Physical Review E.} {\bf 80}(1 Pt 2), 016109.\\
{}[Ronhovde \& Nussinov, 2010] Ronhovde, P.\ \& Nussinov, Z.\ (2010). Local resolution-limit-free Potts model for community detection. {\em Physical Review E.} {\bf 81}(1 Pt 2), 046114.\\
{}[Scott \& Longuet-Higgins, 1990] Scott, G.L.\ \& Longuet-Higgins, H.C.\ (1990). Feature grouping by relocalisation of eigenvectors of the proximity matrix. In {\em Proceedings of British Machine Vision Conference.} 103--108.\\
{}[Shi \& Malik, 2000] Shi, J.\ \& Malik, J.\ (2000). Normalized cuts and image segmentation. {\em IEEE Transactions on Pattern Analysis and Machine Intelligence.} {\bf 22}(8), 888--905.\\
{}[Stringaria {\em et al.}, 2011] Stringaria, C., Cinquinb, A., Cinquinb, O., {\em et al.} (2011). Phasor approach to fluorescence lifetime microscopy distinguishes different metabolic states of germ cells in a live tissue. {\em Proceedings of the National Academy of Sciences of the USA.} {\bf 108}(33), 13582–-13587.\\
{}[Ushakov {\em et al.}, 2011] Ushakov, D.S., Caorsi, V., Ibanez-Garcia, D., {\em et al.} (2011). Response of rigor cross-bridges to stretch detected by fluorescence lifetime imaging microscopy of myosin essential light chain in skeletal muscle fibers. {\em The Journal of biological chemistry.} {\bf 286}(1), 842--850.\\
{}[Verveer {\em et al.}, 2002] Verveer, P.J., Wouters, F.S., Reynolds, A.R., \& Bastiaens, P.\ (2000). Quantitative imaging of lateral ErbB1 receptor signal propagation in the plasma membrane. {\em Science.} {\bf 290}(5496), 567--1570.\\ 
{}[Webb {\em et al.}, 2008] Webb, S.E.D., Roberts, S.K., Needham, S.R., {\em et al.} (2008). Single-molecule imaging and fluorescence lifetime imaging microscopy show different structures for high- and low-affinity epidermal growth factor receptors in A431 Cells. {\em Biophysical Journal.} {\bf 94}(3), 803--819.

\newpage

\myfig{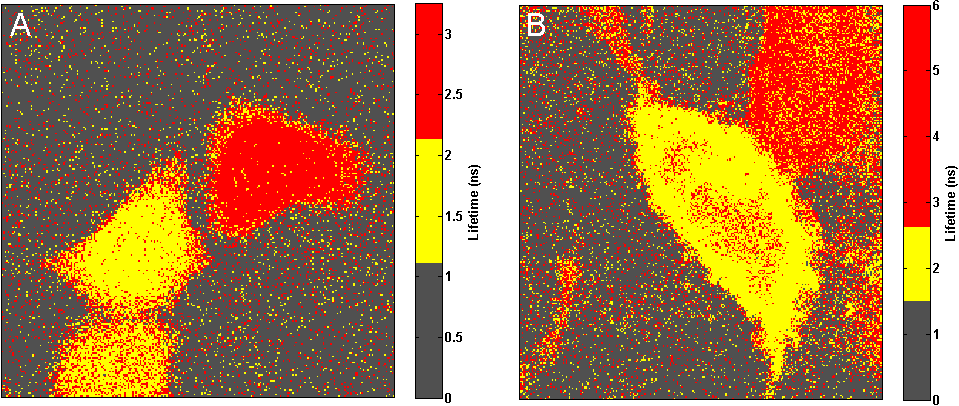}{Fluorescence lifetime imaging microscopy images of live cells transfected with EGFP-RFP fusion F$\ddot{\mathrm{o}}$rster resonance energy transfer (FRET) protein pairs. In both cases, two adjacent cells show distinct donar fluorescent lifetimes (FLTs). One of the cells is a FRET cell and the other cell's acceptor RFP was irreversibly bleached leading to a FLT shift than the former. Images are depicted by carefully adjusting the respective colormaps. False colors represent different segments. These images were used for evaluating performance of the proposed segmentation method.}{fig:fig1}{1\linewidth}{}

\newpage
\myfig{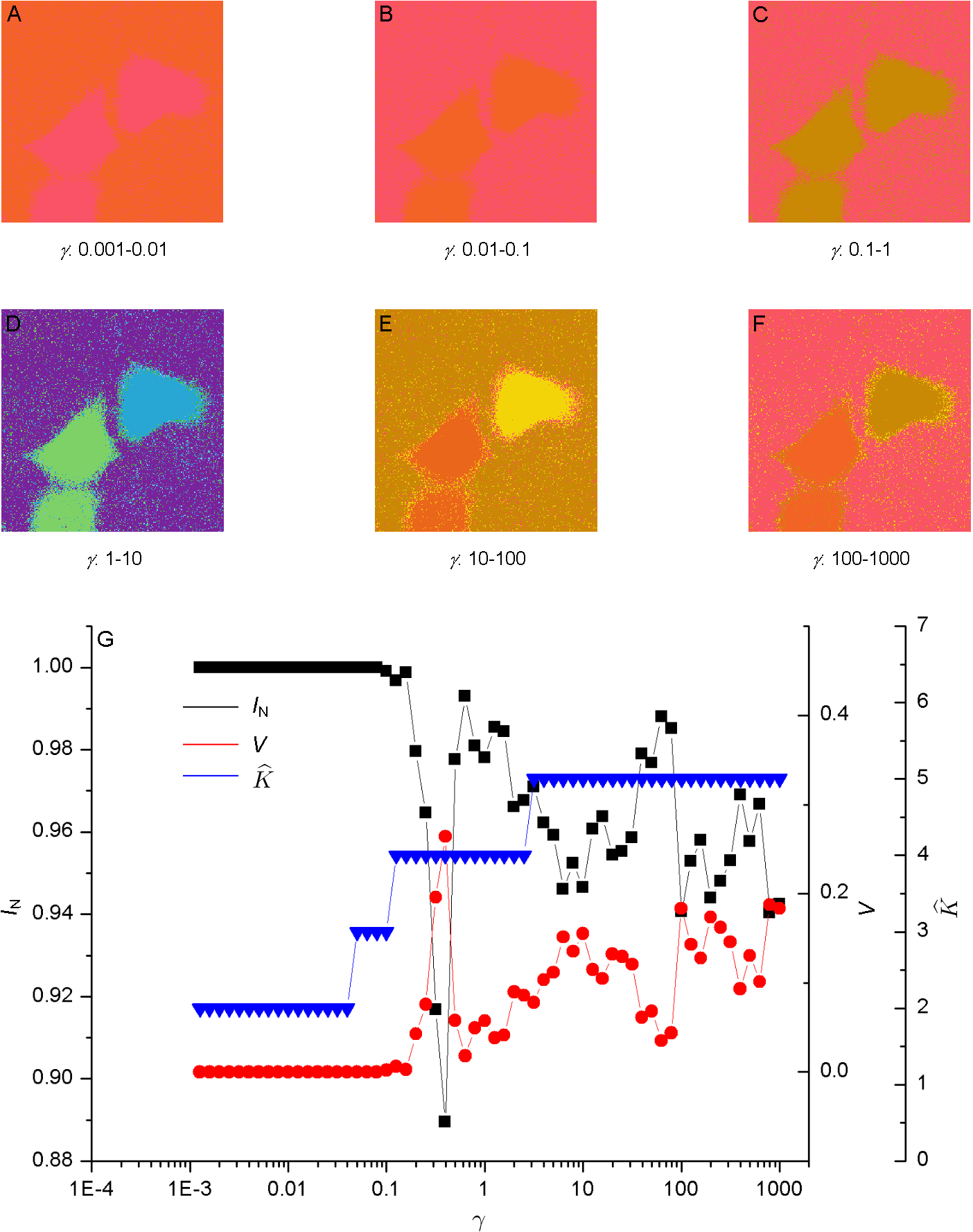}{(A-F) Segmentation of the fluorescence lifetime imaging microscopy image shown in Fig.\ 1A using the multi-resolution community detection (MCD) method for increasing resolution. False colors represent different segments. For resolution parameter $\gamma > 1,$ two major segments, one representing the F$\ddot{\mathrm{o}}$rster resonance energy transfer cell and the other representing the unquenched cell, are clearly visible. (G) Information theoretic overlaps between the replicas of the MCD method, such as their normalized mutual information $I_\mathrm{N}$ and variation of information $V$, together with the number of estimated segments $\widehat{K}$ as a function of $\gamma$. Natural network resolutions correspond to the values of $\gamma$ for which the replicas exhibit extrema and plateau in the average of their information theoretic overlaps.}{fig:fig2}{0.8\linewidth}{}

\newpage
\myfig{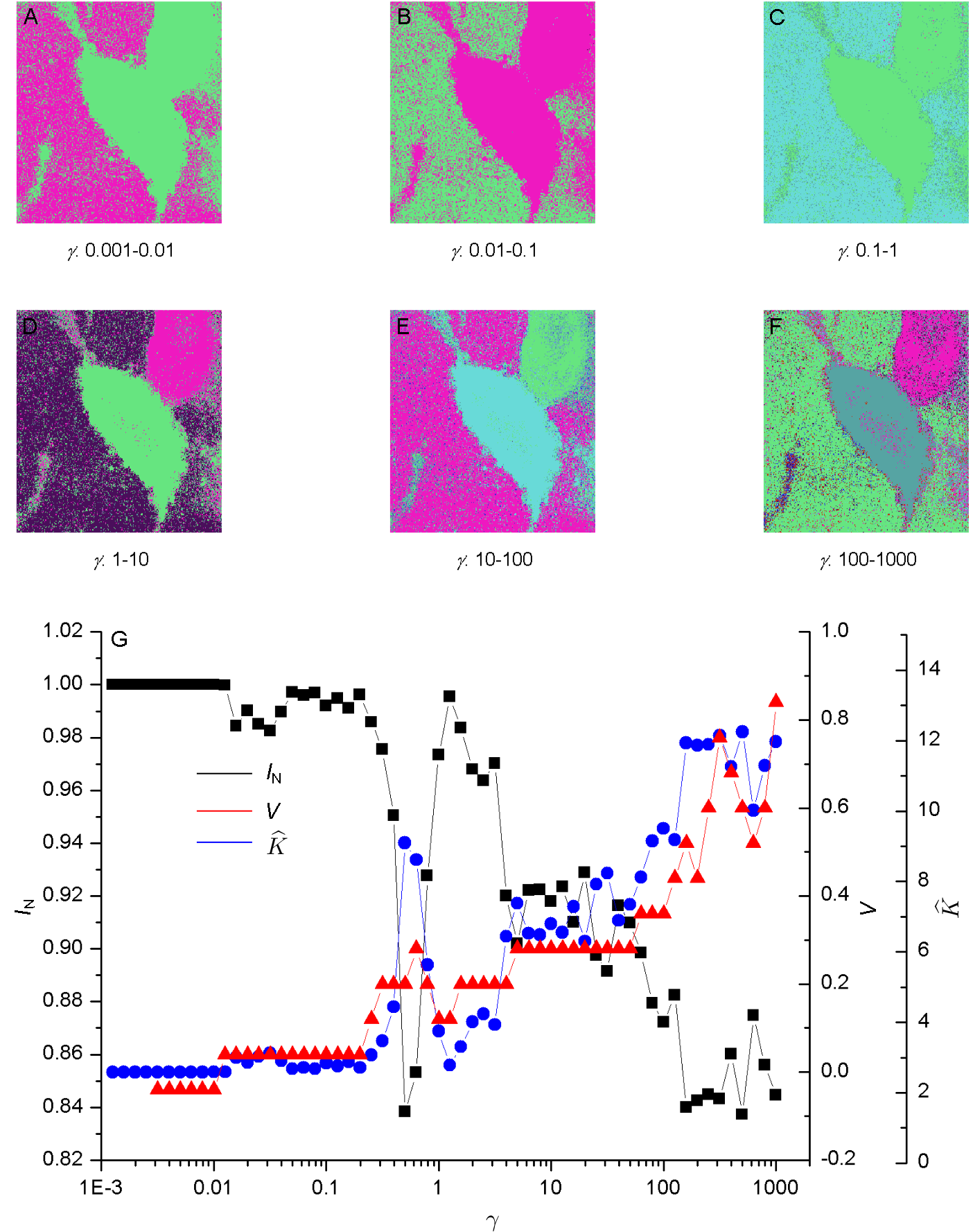}{(A-F) Segmentation of the fluorescence lifetime imaging microscopy image shown in Fig.\ 1B using the multi-resolution community detection (MCD) method for increasing resolution. False colors represent different segments. Identical result was obtained as achieved for Fig.\ 1. (G) Information theoretic overlaps between the replicas of the MCD method, together with the number of estimated segments $\widehat{K},$ as a function of $\gamma$.}{fig:fig3}{0.8\linewidth}{}

\newpage
\myfig{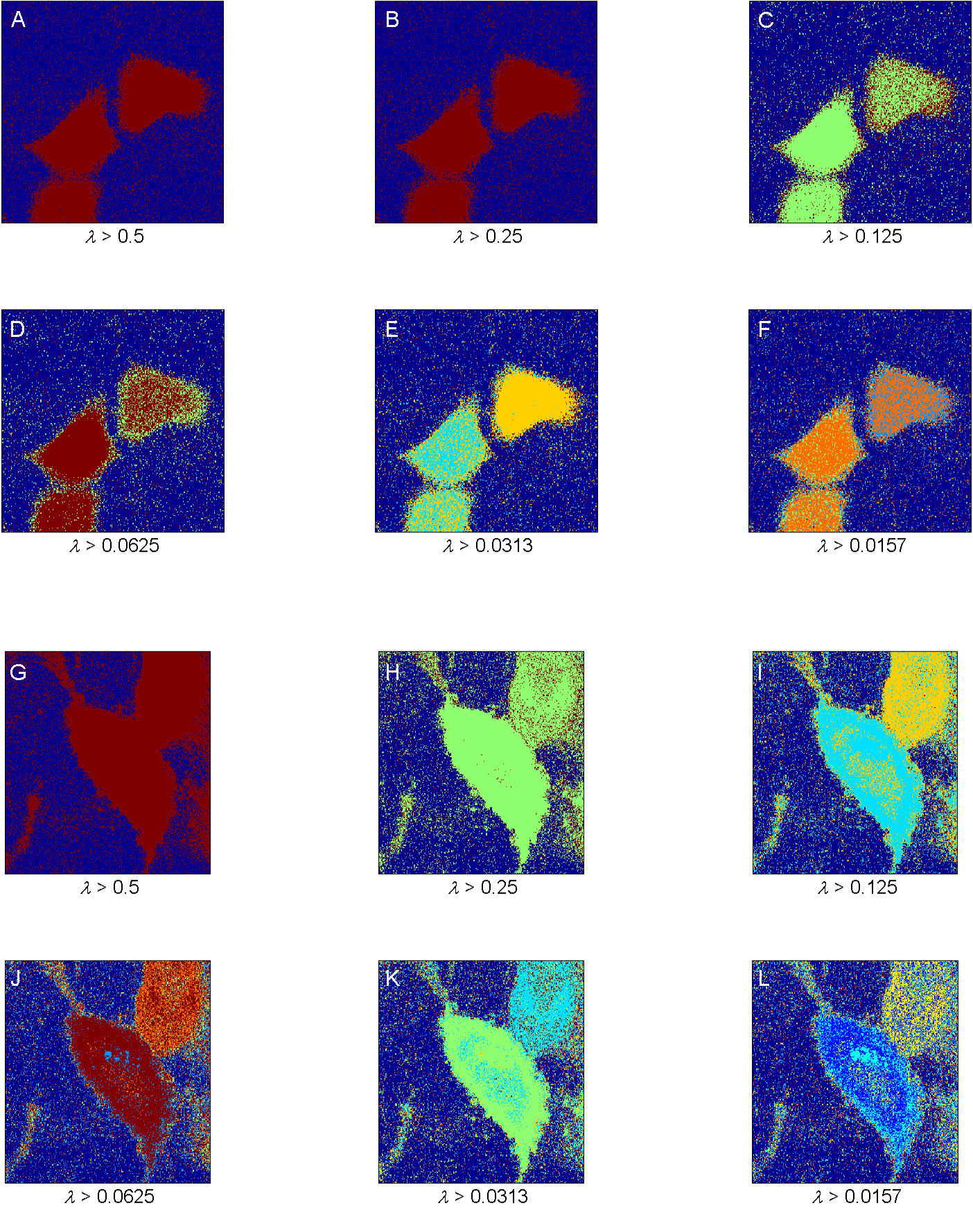}{(A-F) Segmentation of the fluorescence lifetime imaging microscopy (FLIM) image shown in Fig.\ 1A using the spectral clustering method developed by Ng {\em et al.}, 2002 for increasing resolution. False colors represent different segments. Spectral clustering was not able to provide the expected segments, and it introduced high amount of noise in the segmented images. (G-L) Segmentation of the FLIM image shown in Fig.\ 1B using the same spectral segmentation method for increasing resolution. False colors represent different segments. The segmented images show the two cells in a single segment in low resolution (G-H) and in two distinct segments in a slightly higher resolution (I-J). At the resolution limit, similar to the case for the image shown in Fig.\ 1A, the segmented images were introduced with high amount of noise.}{fig:fig4}{0.8\linewidth}{}

\newpage
\myfig{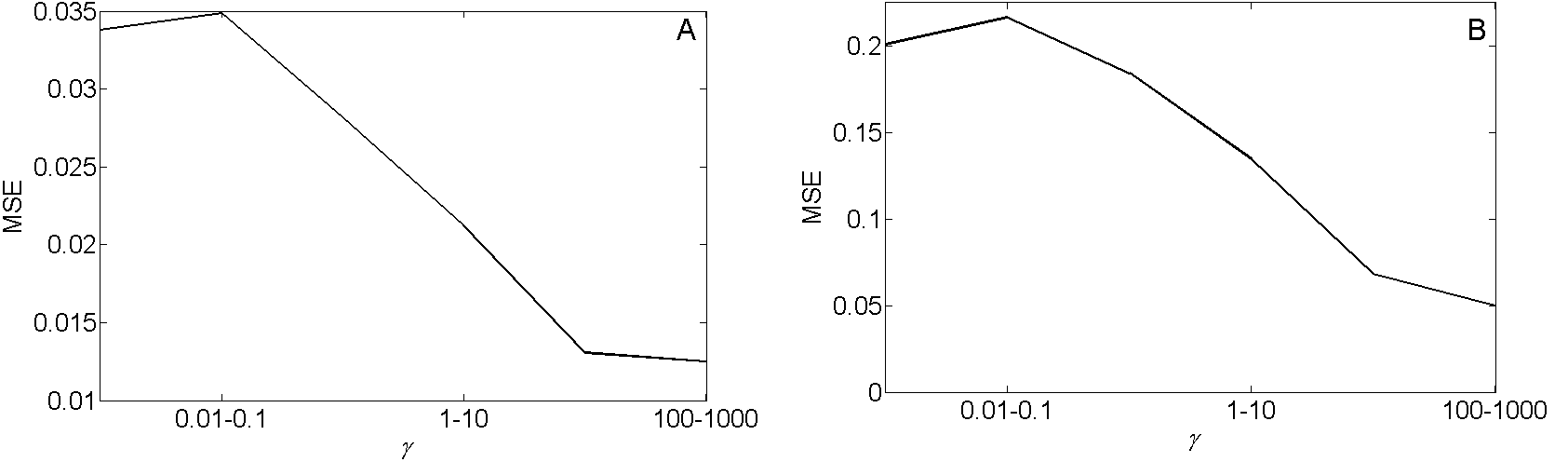}{(A-B) The mean-square error (MSE) in estimating average fluorescence lifetimes of the correct segments using the multi-resolution community detection method for images shown in Fig.\ 1A and Fig.\ 1B, respectively. The MSE consistently decreases with increasing resolution.}{fig:fig5}{1\linewidth}{}

\newpage
\myfig{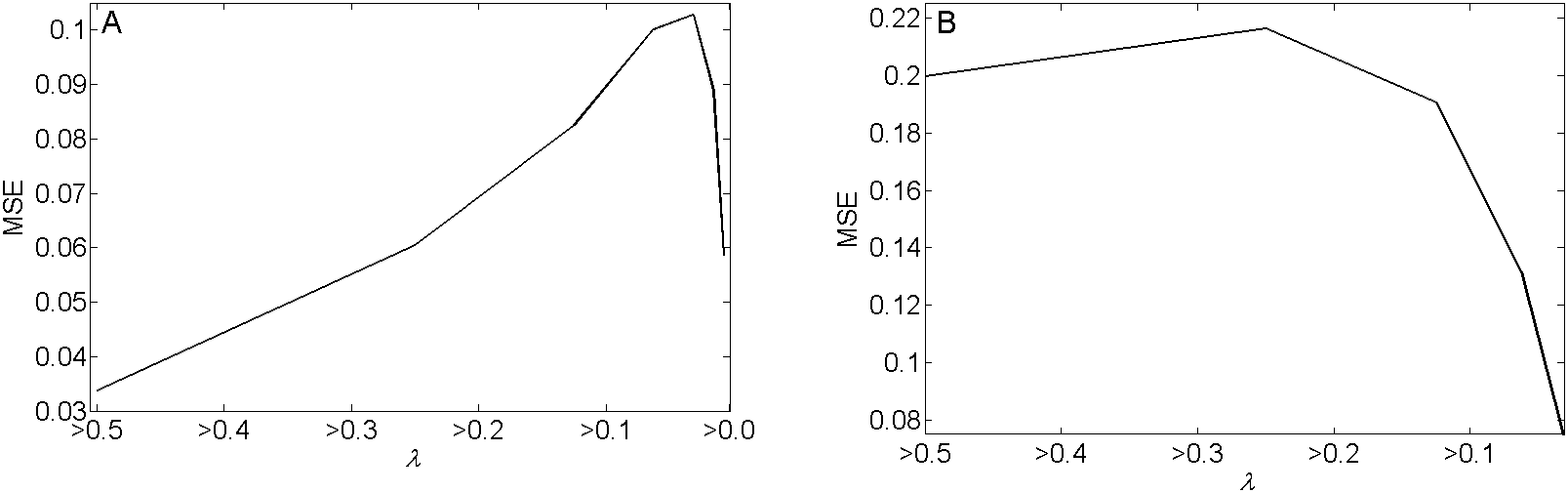}{(A-B) The mean-square error (MSE) in estimating average fluorescence lifetimes of the correct segments using the spectral clustering method developed by Ng {\em et al.} for images shown in Fig.\ 1A and Fig.\ 1B, respectively. The MSE in estimating average FLTs of the correct segments for the image shown in Fig.\ 1A using the spectral clustering method does not consistently decrease with increasing resolution. For Fig.\ 1B, the MSE shows a decrease in increasing the resolution by decreasing $\alpha$ up to 0.0625. Decreasing $\alpha$ below 0.0625 increases the MSE to be very high, and thus, such MSEs are not depicted here for clarity.}{fig:fig6}{1\linewidth}{}

\end{document}